\documentclass[11pt]{iopart}
\usepackage{graphicx}
\usepackage{epsfig,color}
\usepackage{amsmathred,amssymb}
\usepackage{times}
\usepackage{subfigure}

\begin{document}
\title[Exotic spacetimes with superconducting circuits]{\sf \bfseries{One-dimensional sections of exotic spacetimes with superconducting circuits}}

\date{\today}

\author{\sf \bfseries Carlos Sab{\'{i}}n}
\address{Instituto de F\'isica Fundamental, CSIC, Serrano 113-bis
  28006 Madrid, Spain}
  \ead{csl@iff.csic.es}

\date{\today}

\begin{abstract}
We introduce analog quantum simulations of 1+1 dimensional sections of exotic 3+1 dimensional spacetimes, such as Alcubierre warp-drive spacetime, G\"{o}del rotating universe and Kerr highly-rotating black hole metric. Suitable magnetic flux profiles along a SQUID array embedded in a superconducting transmission line allow to generate an effective spatiotemporal dependence in the speed of light, which is able to mimic the corresponding light propagation in a dimensionally-reduced exotic spacetime. In each case, we discuss the technical constraints and the links with possible chronology protection mechanisms and we find the optimal region of parameters for the experimental implementation.
\end{abstract}
\maketitle

\section{Introduction}

Human beings are constantly trying to overcome the constraints imposed by nature. Therefore, it is natural to dream of interstellar journeys, faster-than-light motion or time travel. General Relativity allows to place these dreams into a scientific context, by establishing the physical conditions that these phenomena would require. Along this vein, spacetime metrics enabling apparent faster-than-light motion \cite{alcubierre,morristhorne} -without, of course, superluminal \textit{local} motion-  or travel into the past \cite{godel,kerr,morristhorne2} have been proposed and studied. However, the Einstein equation determines the energy-momentum content of these hypothetical spacetimes and establishes that either they violate the weak energy condition -negative energy densities,- which would require an exotic source of energy, or they rely on provably wrong hypothesis -such as globally rotating universes,- not to mention stability issues related to quantum effects \cite{hawking}. Indeed, Hawking posed a ``chronology protection conjecture" \cite{hawking}, which cannot be fully confirmed or refuted in the absence of a full quantum theory of gravity. 

Among other most practical applications, quantum simulations allow us to make our dreams true, at least up to a certain extent. This is the approach underlying the analog simulations of unseen or impossible physics, for instance, magnetic monopoles \cite{monopole}, tachyonic particles \cite{tachyons} or traversable wormholes \cite{sabinwh, qswh2}. Several modern quantum systems are typically used in these simulations.  One of them is superconducting circuits, where relativistic effects in quantum mechanics and quantum field theory have been analysed, both in a direct or a simulated fashion \cite{casimirwilson,casimirsorin,nationhawking,reviewjohansson,simoneunruh, laura,superluminal}. Indeed, in \cite{sabinwh} we proposed a superconducting-circuit setup for the simulation of a traversable wormhole spacetime. Unlike any alternative classical proposal, in a quantum simulator of an exotic spacetime, quantum effects can give rise to the appearance of analogues of Hawking's chronology protection mechanism \cite{sabinwh}.

In this work, we propose a quantum simulator of several spacetimes with exotic properties, such as Alcubierre warp drive metric \cite{alcubierre} -which can be used for faster-than-light travel,- G\"{o}del rotating universe \cite{godel} and Kerr rotating black holes in the extremal regime \cite{kerr} --which can both support closed timelike curves-. In particular, we do not consider the full 3+1 dimensional spacetimes, but only  1+1 dimensional slices of them. Then, the spacetime metric can be encoded into the spatiotemporal dependence of the speed of light. Effective profiles for the variation in space and time of the speed of propagation of the electromagnetic field can be generated in several superconducting circuit setups. In particular, here we consider an SQUID array embedded into a microwave transmission line, where the speed of light is given by the inductance of the SQUIDs and can be modulated through an external magnetic flux. In each of the cases considered, we find the required profiles in space and time of the magnetic flux. Theoretical and technological limitations appear, thus constraining both the parameters of the spacetime that we are able to simulate and the values of the fluxes that we need to use in order to implement the simulator. In all the cases, we perform a thorough analysis of the optimal parameter regimes and their physical implications. 

\section{1+1 spacetime sections}

We will start by discussing our approach to 1+1 dimensional spacetimes. In 1+1 D, the Klein-Gordon equation of a scalar field -such as a one-dimensional electromagnetic field- possesses conformal invariance. We can leverage this feature to consider a family of \textit{dimensionally reduced} spacetimes of the form
\begin{equation}\label{eq:spacetime}
ds^2=-c^2(r,t)dt^2+dr^2,
\end{equation}
where we have a Minkowski-like spacetime with an effective speed of light depending on the spatiotemporal coordinates $r$ and $t$, and we simply ignore the other coordinates. We will bring all the spacetimes of interest into this form, so the spacetime features will be encoded into an effective speed of light profile. 

If we treat Eq. (\ref{eq:spacetime}) as the metric of an actual 1+1 D spacetime, we find that the Ricci tensor $R_{\mu\nu}$  is diagonal, with components
\begin{equation}\label{eq:ricci}
R_{rr}=-\frac{c''(r,t)}{c(r,t)},\,\, R_{tt}=c''(r,t)c(r,t),
\end{equation}
where the $'$ denotes derivative with respect to the spatial coordinate $r$. The scalar curvature $R$ turns out to be 
\begin{equation}\label{eq:scalarcurvatrue}
R=-2\frac{c''(r,t)}{c(r,t)}.
\end{equation}
Finally, the Einstein tensor is null and therefore it satisfies the vacuum Einstein equation. Indeed, this is actually the case of any 1+1 D metric \cite{collas}, which of course questions the meaning of General Relativity in 1+1 D. However, we remark that we are not interested in Eq. (\ref{eq:spacetime}) as a spacetime metric \textit{per se}, but only as a section of a full 3+1 D spacetime, which will of course have non-trivial energy-momentum tensor, even in the $\{r,t \}$ sector.

\section{Effective speed of light}

An effective speed of light depending on space and/or time can be generated in several 1+1 D superconducting circuit setups. In this work we consider a dc-SQUID array embedded in an open transmission line \cite{casimirsorin,array, array2,array3}. The speed of propagation along the transmission line is given by
\begin{equation}\label{eq:squidspeed}
c^2(\phi_{ext})=c_0^2|\cos\frac{\pi\phi_{ext}}{\phi_0}|
\end{equation}
where we are denoting $c_0$ as the speed of light in the absence of external flux $c_0^2=c^2(\phi_{ext}=0)=1/(L_s(\phi_{ext=0})C_s)$, which is given by the inductance $L_s$ and capacitance $C_s$ of the SQUIDs. Here $\phi_0=h/(2\, e)$ is the flux quantum, and $\phi_{ext}$ is the external magnetic flux threading the SQUID. We can split the external magnetic flux into two contributions
\begin{equation}\label{eq:acdc}
\phi_{ext}(r,t)=\phi_{ext}^{DC}+\phi_{ext}^{AC}(r,t).
\end{equation}
If we bias all the SQUIDs with only the constant DC term, this has the effect of reducing the effective speed of light along the whole SQUID array:
\begin{equation}\label{eq:squidspeeddc}
c^2(\phi_{ext}^{DC})=c_0^2|\cos\frac{\pi\phi_{ext}^{DC}}{\phi_0}|.
\end{equation}
If we now add a spacetime dependent AC signal and use a standard trigonometric identity, we have:
\begin{equation}\label{eq:squidspeedacddc}
c^2(\phi_{ext})=c^2(\phi_{ext}^{DC})(|\cos\frac{\pi\phi_{ext}^{AC}}{\phi_0}-\tan\frac{\pi\phi_{ext}^{DC}}{\phi_0}\sin{\frac{\pi\phi_{ext}^{AC}}{\phi_0}}|).
\end{equation}
Then, we have a new contribution to the effective speed of light, which we can denote by $\widetilde{c}^2(\phi_{ext})$:
\begin{equation}\label{eq:squidspeedac}
\widetilde{c}^2(\phi_{ext})= |\cos\frac{\pi\phi_{ext}^{AC}}{\phi_0}-\tan\frac{\pi\phi_{ext}^{DC}}{\phi_0}\sin{\frac{\pi\phi_{ext}^{AC}}{\phi_0}}|.
\end{equation}
Using the trigonometric identity $\cos A-\tan B\sin A=\sec B\cos{A+B}$, we can rewrite it as:
\begin{equation}\label{eq:squidspeedac2}
\widetilde{c}^2(\phi_{ext})=|\sec\frac{\pi\phi_{ext}^{DC}}{\phi_0}||\cos\frac{\pi\phi_{ext}}{\phi_0}|,
\end{equation}
and restricting ourselves to the region $ [-\pi/2,\pi/2] $ for both the DC part $\pi\phi_{ext}^{DC}/\phi_0$ and the total flux $\pi\phi_{ext}/\phi_0$, we can get rid of the absolute values. 
Then, finally we have that the effective speed of light is:
\begin{equation}\label{eq:finalspeedsquids}
c^2(\phi_{ext})=c^2(\phi_{ext}^{DC})\widetilde{c}^2(\phi_{ext}),
\end{equation}
where $c^2(\phi_{ext}^{DC})$ is given by Eq. (\ref{eq:squidspeeddc}) and $\widetilde{c}^2(\phi_{ext})$ by Eq.(\ref{eq:squidspeedac2}).

Now, if we want to use this to simulate the spacetime family in Eq. (\ref{eq:spacetime}), we can split the effective speed of light in the spacetime into two parts, namely a constant one $c^2$ and a spacetime-dependent one $\widetilde{c}^2(r,t)$, such that:
\begin{equation}\label{eq:reallightspeed}
c^2(r,t)=c^2\widetilde{c}^2(r,t).
\end{equation}
Then the task is to relate Eq. (\ref{eq:reallightspeed}) to Eq. (\ref{eq:finalspeedsquids}). In the first part, we simply choose a value of the DC flux, which sets the value $c^2$ of the speed of light in a simulated flat spacetime, which might be significantly smaller than the value of the speed of the light in vacuum and than the speed of light in a transmission line in the absence of any external bias:
\begin{equation}\label{eq:rosetta1}
c^2=c_0^2|\cos\frac{\pi\phi_{ext}^{DC}}{\phi_0}|.
\end{equation}
In the second part, the AC  part of the flux allows us to simulate a spatiotemporal profile for the speed of light:
\begin{equation}\label{eq:rosetta2}
\widetilde{c}^2(r,t)=\sec\frac{\pi\phi_{ext}^{DC}}{\phi_0}\cos\frac{\pi\phi_{ext}}{\phi_0}.
\end{equation}
Therefore, for a given $\widetilde{c}^2(r,t)$ of interest, the task is to choose a flux profile such that:
\begin{equation}
\frac{\pi\phi_{ext}^{AC}(r,t)}{\phi_0}=\arccos[\widetilde{c}^2(r,t)\cos\frac{\pi\phi_{ext}^{DC}}{\phi_0}]-\frac{\pi\phi_{ext}^{DC}}{\phi_0}.
\end{equation}

Now, we will apply these results to several exotic spacetimes of interest.

\section{Warp drives}

In \cite{alcubierre}, Alcubierre proposed a simple spacetime metric which allows to travel "faster than light", in the sense of faster than the speed of light in the vacuum of a flat spacetime. He introduced a `bubble" of spacetime moving along the $x$ axis with velocity $v_s$. This produces the following metric:
 \begin{equation} d s^2 \,=\, - c^2d t^2 \,+\, \left(
\rule{0mm}{4mm} d x - v_s \, f \left( r_s \right) \, d t \right)^2
\,+\, d y^2 \,+\, d z^2 \;\; .  \label{final_metric} \end{equation}
where
 \[ v_s \left( t \right) \,=\, \frac{d x_s \left( t
\right) }{d t} \;\; ,  r_s \left( t \right) \,=\, \left[
\left( \rule{0mm}{4mm} x - x_s \left( t \right) \right)^2 \,+\, y^2
\,+\, z^2 \, \right]^{1/2} \; ,  \]
and  \,$f$\, is the function:  \begin{equation} f \left(
r_s \right) \,=\, \frac{\tanh \left( \rule{0mm}{4mm} \sigma \, \left(
r_s \,+\, R \right) \right) \,-\, \tanh \left( \rule{0mm}{4mm} \sigma
\, \left( r_s \,-\, R \right) \right) }{ \rule{0mm}{5mm} 2 \, \tanh
\left( \sigma \, R \right)} \;\; , \end{equation}

\noindent with \,$R\,>\,0$\, and \,$\sigma\,>\,0$\, arbitrary
parameters.  Notice that for large \,$\sigma$\:  
\begin{equation}
\lim_{\sigma \rightarrow \infty} \, f \left( r_s \right) \;\;=\;\; \left\{ \begin{array}{l} 1 \hspace{5mm}  {\rm for}
\hspace{4mm} r_s \,\in\, \left[ -R,\,R \right] \;\; , \\ 0 \hspace{5mm}
{\rm otherwise} \;\; .  \end{array} \right.  \end{equation}
In this way, the spacetime can be be split into two parts: inside the moving bubble of radius $R$ we have 
\begin{equation} d s^2 \,=\, - c^2d t^2 \,+\, \left(
\rule{0mm}{4mm} d x - v_s  \, d t \right)^2
\,+\, d y^2 \,+\, d z^2 \;\; .  \label{final_metric_ins} \end{equation}
while outside the bubble we have a flat spacetime.

In principle, nothing prevents us from considering $v_s>c$. Therefore the bubble is moving superluminally in this sense. However, this does not mean that the bubble is traveling faster than the \textit{local} speed of light, which is of course forbidden in General Relativity. Indeed, by determining the null geodesic $ds^2=0$, we see that the speed of light inside the bubble is 
\begin{equation}\label{speedoflightins}
c_s= c+v_s.
\end{equation}
Therefore $c_s>v_s$ always. 

With an SQUID array, we can simulate a 1+1 D spacetime where the speed of light mimics the speed of light in Alcubierre spacetime. Note that this does not mean that we are simulating the Alcubierre metric, nor the Alcubierre metric under any coordinate transformation. We are simulating an spacetime of the form in Eq. (\ref{eq:spacetime}) in which the speed of light profile is the same as the speed of light profile in Alcubierre spacetime, in the same coordinates.  We can refer to it as  Alcubierre-like spacetime, in which the speed of light is indistinguishable from the speed of light in Alcubierre spacetime Therefore, using the results of the previous section, the task is to produce a profile $\widetilde{c}^2(r,t)$ such that:
\begin{equation}\label{eq:profalc}
\widetilde{c}^2(r,t)=\left(1+\frac{v_s}{c}\right)^2,
\end{equation} 
inside a moving bubble of a giving length, while keeping  a constant value $c$ outside. 

Inserting Eq. (\ref{eq:profalc}) in Eq. (\ref{eq:finalspeedsquids}), we see that, since the argument of $\arccos$ is of course restricted to $[-1,1]$, this translates into a restriction for the DC and AC fluxes that we can consider. Therefore, it is interesting to explore whether is possible to actually achieve a superluminal $v_s$. 

\begin{figure}[t!]\centering
\includegraphics[width=0.9\textwidth]{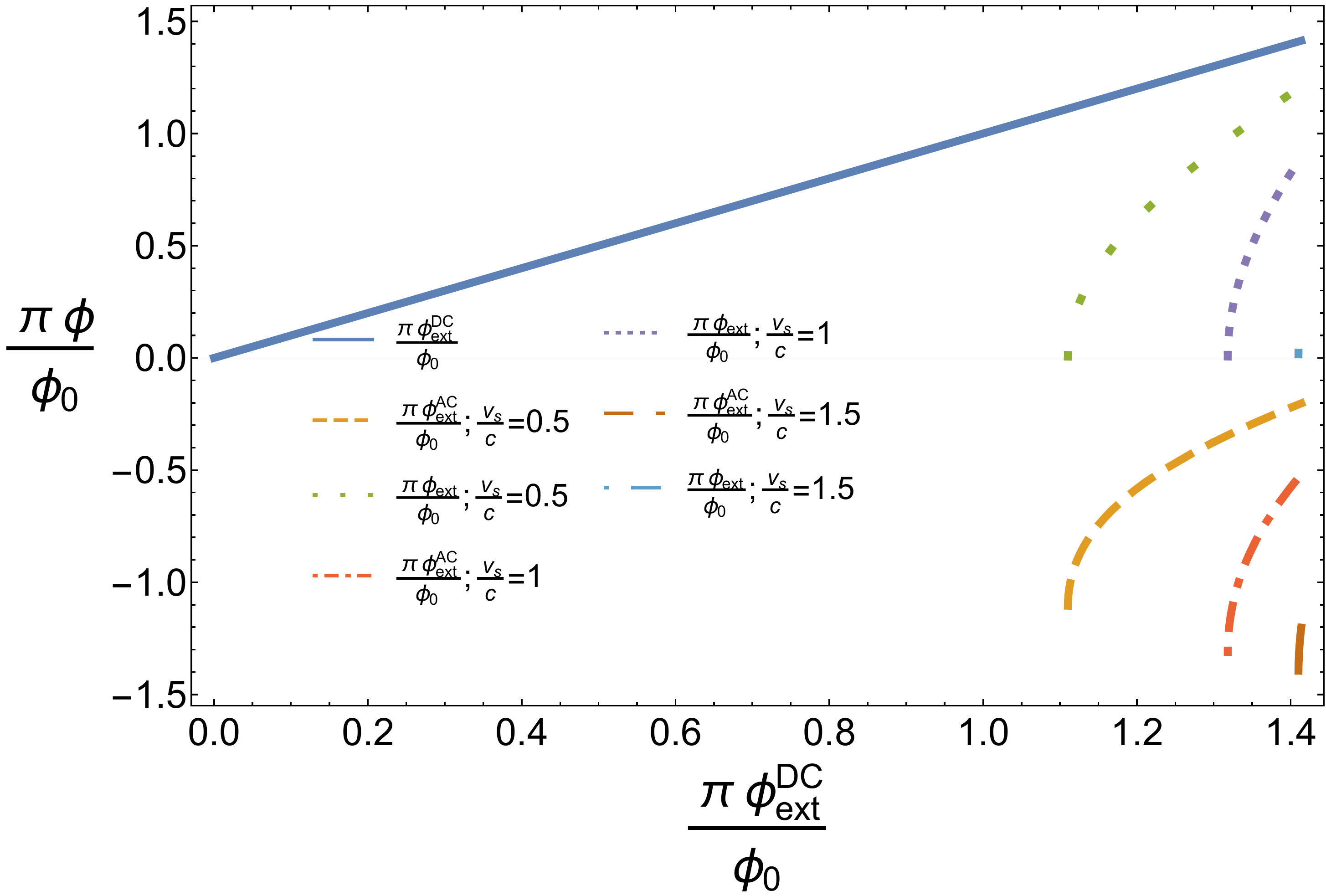}  
\caption{Magnetic fluxes $\frac{\pi\phi_{ext}^{DC}}{\phi_0}$, $\frac{\pi\phi_{ext}^{AC}}{\phi_0}$ and $\frac{\pi\phi_{ext}}{\phi_0}$ for different simulated velocities of the moving spacetime bubble in Alcubierre spacetime $\frac{v_s}{c}$, ranging from subluminal $\frac{v_s}{c}=0.5$ to superluminal $\frac{v_s}{c}=1.5$.}
   \label{Fig1}
  \end{figure}

In Fig. (\ref{Fig1}), we see that not all the values of the fluxes are allowed, and that the restrictions are more constraining as we consider increasing constant velocities $v_s$. Moreover, we have the additional technical difficulty already discussed in \cite{sabinwh}: the value of the flux cannot be too close to $\pi\phi_0/2$ along all the array, otherwise the impedance of the array would trigger quantum fluctuations of the superconducting phase, breaking down the approximations which led to Eq. (\ref{eq:squidspeeddc}). However, we see in Fig. (\ref{Fig1}) that, even in the case $v_s/c=1.5$, it is possible to find some allowed values of the flux. For instance, we can choose for the DC part $\pi\phi_{ext}^{DC}/\phi_0\simeq -0.44 \pi$, -in the edge of the low-impedance approximation- and a very similar value with opposite sign for the AC part. Then the AC part should move at the velocity $v_s$ along the array. Note that $v_s$ is superluminal only with respect to the background velocity, which has been reduced to approximately $0.4$ its typical value, by means of the chosen $DC$ flux. Then, the  AC flux should  move at approximately $0.6 c_0$, $c_0$ being the value of the speed of light in the absence of any flux. 
Note also that quantum fluctuations of the superconducting phase could be interpreted as an analogue of Hawking's chronology protection mechanism trying to prevent us from building up an analogue exotic spacetime with causality issues.

\section{G\"{o}del spacetimes}

In \cite{godel}, G\"{o}del proposed an interesting metric allowing time travel, which was the solution of the Einstein equation in a rotating spacetime with cosmological constant. 
The resulting line element in cylindrical coordinates reads \cite{godelcil}
\begin{eqnarray}
ds^2&=&c^2dt^2-\frac{dr^2}{1+\left(\frac{r}{2a}\right)^2}-r^2\left(1-\left(\frac{r}{2a}\right)^2\right)d\phi^2\nonumber\\
&-&dz^2+2r^2\frac{c}{\sqrt{2}a}dt\,d\phi\;. 
\label{metricgodel}
\end{eqnarray}
where $a$ is the G\"{o}del parameter. 

Reducing to 1+1 D $\{t,r\}$ and finding the null geodesics $ds^2=0$, we find that the speed of light profile is:
\begin{equation}
\label{eq:profgod}
\widetilde{c}^2(r,t)=1+\left(\frac{r}{2 a}\right)^2,
\end{equation} 
\begin{figure}[t!]\centering
\includegraphics[width=0.9\textwidth]{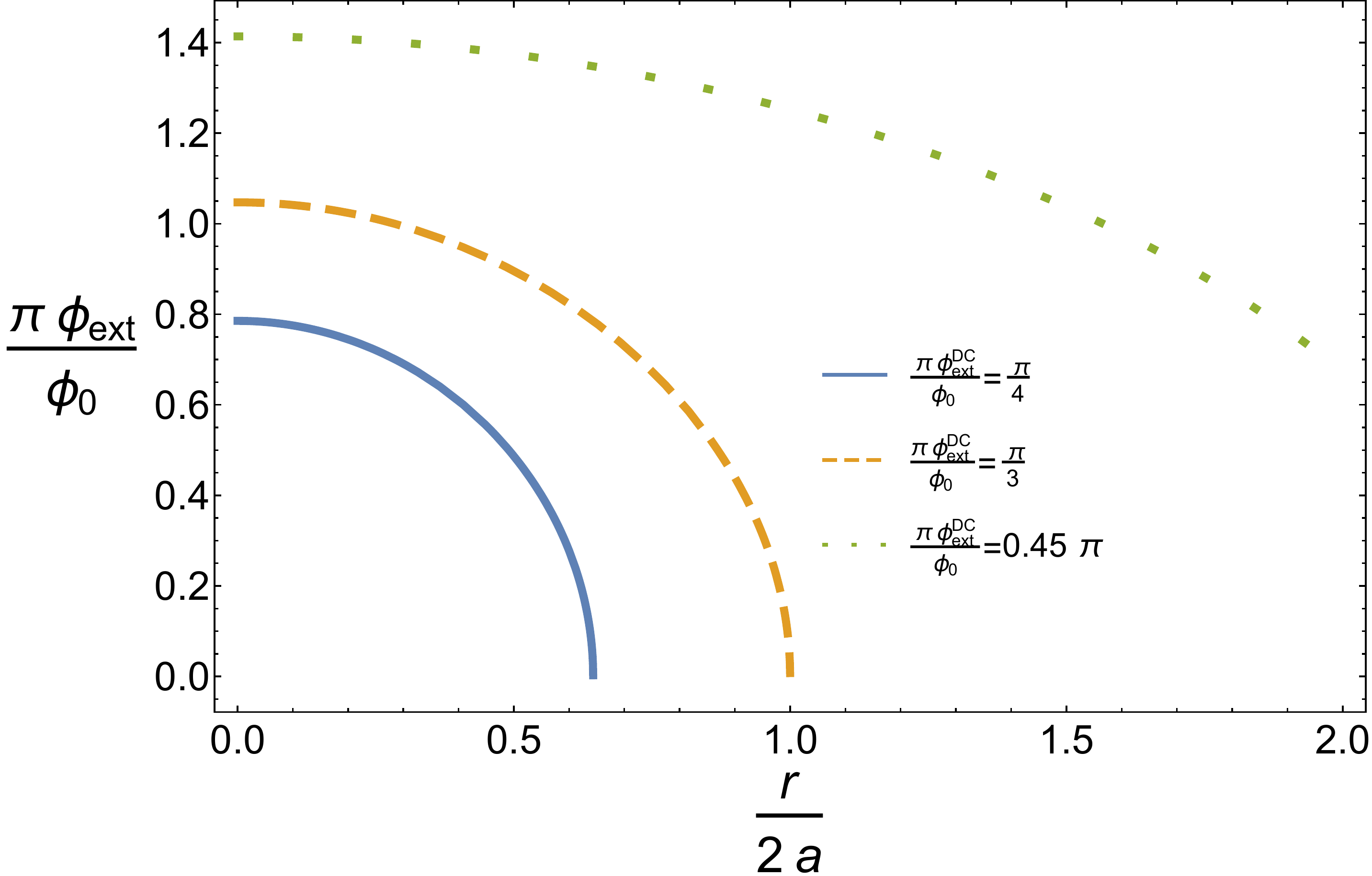}  
\caption{Magnetic flux $\frac{\pi\phi_{ext}}{\phi_0}$ vs. radial coordinate in G \"{o}del spacetime $\frac{r}{2 a}$ for different values of the DC part of the flux $\frac{\pi\phi_{ext}^{DC}}{\phi_0}$. The maximum size of the simulated spacetime depends on this value.}
   \label{Fig2}
  \end{figure} 
which in this case is independent of $t$. Inserting Eq. (\ref{eq:profgod}) in Eq. (\ref{eq:finalspeedsquids}), we see that now similar restrictions as in the Alcubierre case would appear, although in this case the constraints are related to the size of the simulated universe with respect to the G\"{o}del parameter. 

In Fig. (\ref{Fig2}), we see that, depending on the chosen value of the DC flux, we are able to simulate spacetimes with maximum radius lying within or outside the G\"{o}del parameter. The latter possibility would be needed in order to use this effective spacetime to simulate time-traveling scenarios, since closed timelike curves would go from inside to outside a cylinder with radius $2a$ \cite{pfarr}.

\section{Extreme Kerr black holes}

In Boyer-Lindquist coordinates and $c=G=1$ units, the line element of the Kerr rotating black hole metric reads~\cite{kerr,bambi}
\begin{eqnarray}
\hspace{-0.5cm}
ds^2 &=& - \left(1 - \frac{2 M r}{\Sigma}\right) dt^2 
- \frac{4 M a r \sin^2\theta}{\Sigma} dt d\phi 
+ \frac{\Sigma}{\Delta} dr^2 \nonumber\\
&& + \Sigma d\theta^2 
+ \left(r^2 + a^2 + \frac{2 M a^2 r \sin^2\theta}{\Sigma} \right) 
\sin^2\theta d\phi^2 \, ,
\end{eqnarray}
where $a = J/M$, $\Sigma = r^2 + a^2 \cos^2 \theta$, and $\Delta = r^2 - 2 M r + a^2$, $J$ and $M$ being the spin and mass of the black hole, respectively. In particular, we will consider the so-called extreme regime, where $a=M$ and the inner and event horizons coincide at $r=M$. Then, 
\begin{equation}\label{eq:kerrext}
\Sigma = r^2 + M^2 \cos^2 \theta,\,\, \Delta=(M-r)^2.
\end{equation}
Following the same procedure as in the two cases above, and setting in this case $\phi_{ext}^{DC}=0$ -since in this case, we have checked that the conclusions below do not change if we vary the value of the DC flux- we find 
\begin{equation}\label{eq:profkerr}
\widetilde{c}^2(r,t)=\left(1-\frac{2 M r}{r^2 + M^2 \cos^2 \theta}\right)\frac{(M-r)^2}{r^2 + M^2 \cos^2 \theta},
\end{equation} 
which depends on $\theta$ -in our case, $\theta$ is not a coordinate, but a parameter- and again, does not depend on $t$. 
\begin{figure}[t!]\centering
\includegraphics[width=0.9\textwidth]{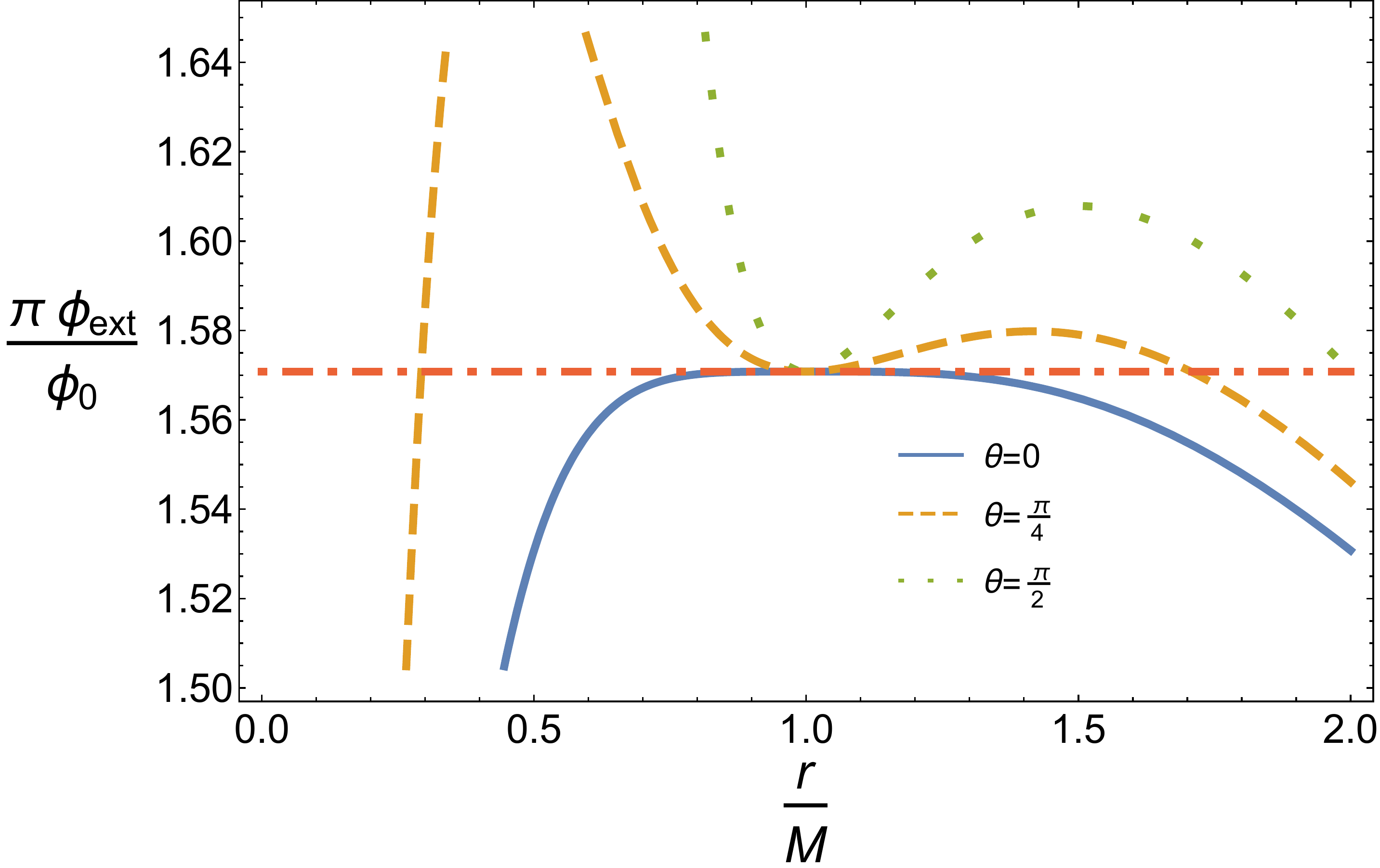}  
\caption{Magnetic flux $\frac{\pi\phi_{ext}}{\phi_0}$ vs. radial coordinate in extreme Kerr spacetime $\frac{r}{M}$ for different values of $\theta$, and zero DC fluxes in all cases. The red dash-dotted horizontal line benchmarks the value $\frac{\pi\phi_{ext}}{\phi_0}=\frac{\pi}{2}$Only in the case $\theta=0$ we find a large continuous region of the spacetime which might be simulated.}
   \label{Fig3}
  \end{figure} 

In Fig. (\ref{Fig3}), we see that for $\theta=\pi/2$ the simulation is not possible since it would require magnetic fluxes above the threshold value $\phi_{ext}/\phi_0=\pi/2$. However, choosing other spacetime slices $\theta=0$, $\pi/4$, we are able to simulate both the interior and exterior parts of the Kerr metric. In the case of $\pi/4$ there is still some forbidden region. However, for $\theta=0$, the only problem is the appearance of a region near the horizon where the flux is exactly $\pi/2$. As in the case of the traversable wormhole spacetime \cite{sabinwh}, the idea would be to choose values of $M$ so that the corresponding values of $r/M$ correspond to a very small region in the laboratory, ideally to only a single SQUID.

\section{Conclusions}

In conclusion, we have shown that suitable spatiotemporal profiles of external magnetic fluxes threading an SQUID array embedded into a microwave transmission line, can generate an effective spacetime dependence in the speed of propagation of the electromagnetic field along  the line, which is able to mimic the propagation of light along one-dimensional slices of several exotic spacetimes. We have considered several paradigmatic scenarios. In the case of Alcubierre warp-drive and G\"{o}del spacetimes we use both DC and AC magnetic fluxes. The former is used in order to reduce the background speed of light in the simulated spacetime, so the latter can be used to simulate superluminality with respect to such speed. In the Alcubierre spacetime, the other part of the flux takes non-zero values only in a small moving region, simulating a moving spacetime bubble, which, in particular can reach superluminal velocities -in the sense of velocities larger than the background velocity, which has been significantly reduced with respect to the one in the absence of any flux. In the G\"{o}del case, the flux profile does not depend on time, but it does depend on the spatial coordinate. Mathematical and technical limitations constrain the maximum velocities that can be simulated in Alcubierre spacetime, as well as the size of the simulated G\"{o}del universe. Finally, we consider the case of an extreme Kerr black hole, where a DC flux is not needed, but forbidden regions appear for values of $\theta\neq0$. In the case of $\theta=0$, the flux must take the value $\phi/\phi_0=\pi/2$ in a neighbourhood of the Kerr horizon. This is the same scenario as in the throat of a traversable wormhole \cite{sabinwh}. Such a value of the flux on a large region of the array would make the impedance of the array grow too high, triggering quantum fluctuations of the superconducting phase and breaking down the approximations of the model. Even  if we restrict the high value of the flux to a small region in the laboratory, it is an open question whether the large impedance in that region would prevent the simulation, a mechanism which might be linked to Hawking's chronology protection conjecture, as already suggested in \cite{sabinwh}. Actually, this could be the most interesting point of the experiment since it can only be addressed in a fully quantum simulator. By coupling superconducting qubits to the SQUID array and simulating their motion \cite{simoneunruh, laura,superluminal}, we could analyze the projection of closed timelike curves in the radial direction of Kerr and G\"{o}del spacetimes. A detailed analysis lies beyond the scope of the current work.

\section*{Acknowledegements}
Financial support from Fundaci{\'o}n General CSIC (Programa ComFuturo) is acknowledged.

\section*{Additional information} 
\subsection*{Competing financial interests}
The author declares no competing financial interests.

\end{document}